\documentclass[superscriptaddress,prd,aps,showpacs,secnumarabic]{revtex4-1}
\usepackage{graphicx,color}
\usepackage{rotating}
\def\Dot{\!\cdot\!}
\def\al{\alpha}

\def\part{\partial}

\begin{document}

\title{Self Coupling of the Higgs boson in the processes 
$p\,p\,\rightarrow\,ZHHH+X$ and $p\,p\,\rightarrow\,WHHH+X$}

\author{Duane A. Dicus}\email{dicus@physics.utexas.edu}\affiliation{Department of Physics, University of Texas, Austin, TX 78712, USA}
\author{Chung Kao}\email{kao@physics.ou.edu}\affiliation{Homer L. Dodge Department of Physics, University of Oklahoma, Norman, OK 73019, USA}
\author{Wayne W. Repko}\email{repko@pa.msu.edu} \affiliation{Department of Physics and Astronomy, Michigan State University, East Lansing, MI 48824, USA}
\date{\today}
\begin{abstract}
To gain some sense about the likelihood of measuring the Higgs boson quartic coupling, we calculate the contribution to the triple Higgs production cross section from the subprocesses $q\bar{q}\to ZHHH$ and $q\bar{q}'\to WHHH$. Our results illustrate that determining this coupling, or even providing experimental evidence that it exists, will be very difficult.   
\end{abstract}

\maketitle

%------------------------
% Make Title Page
%------------------------
\maketitle

%=======================================================================
%   BEGIN MAIN TEXT
%=======================================================================
\newpage

% Section 1
\section{Introduction}

The Standard Model (SM) has been very successful in explaining almost
all experimental data to date, culminating in the discovery of
the long awaited Higgs boson at the CERN Large Hadron Collider 
(LHC)~\cite{Aad:2012tfa,Chatrchyan:2012xdj}.
The most important experimental goals of Run 2 at the Large Hadron 
Collider are the investigation of Higgs properties and the search for 
new physics beyond the Standard Model.

Thus far the results from the LHC indicate that the couplings of
the Higgs boson to other particles are consistent with the
Standard Model. However the ultimate test as to whether this particle
is the SM Higgs boson will be the trilinear Higgs coupling that
appears in Higgs pair production and the quartic Higgs coupling 
that shows up in triple Higgs production.

The self interaction of the Higgs field, $H$, is
\begin{equation}\label{V(H)}
V(H)\,=\,\lambda\,v^2\,H^2\,+\,\kappa_3\lambda\,v\,H^3\,+\,\frac{1}{4}\kappa_4\,\lambda\,H^4
\end{equation}
where $\lambda\,v^2\,=\,\frac{1}{2}m_H^2$ and $v$ is the vacuum expectation value given by the $Z$ mass, $M_Z$, the weak mixing angle $\theta_W$, and the fine structure constant $\al$ as $v\,=\,M_Z\,\cos\theta_W\sin\theta_W/\sqrt{\pi\alpha}$. $\kappa_3$ and $\kappa_4$ are one in the standard model; these are what we would like to measure.

\begin{figure}[h]
\begin{minipage}[t]{0.30\textwidth}
\centering\includegraphics[height=1.25in]{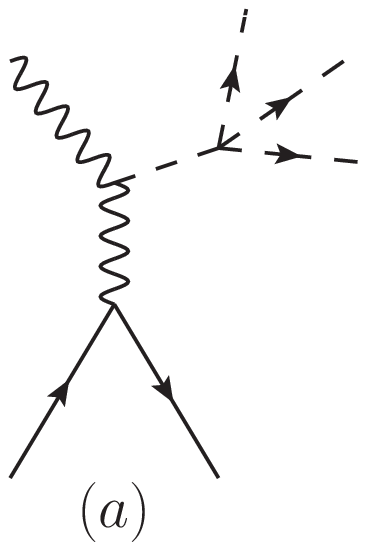}
\end{minipage}%
\begin{minipage}[t]{0.05\textwidth}
\hfil
\end{minipage}%
\begin{minipage}[t]{0.30\textwidth}
\centering\includegraphics[height=1.25in]{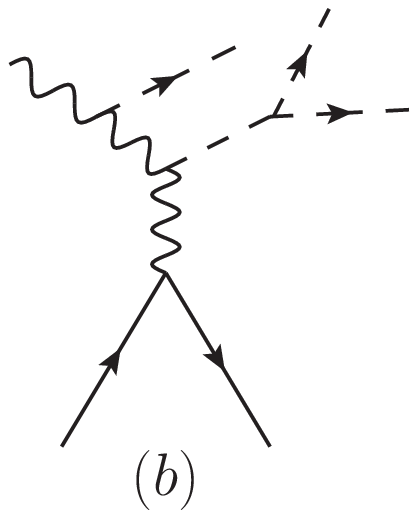}
\end{minipage}%
\begin{minipage}[t]{0.05\textwidth}
\hfil
\end{minipage}%
\begin{minipage}[t]{0.30\textwidth}
\centering\includegraphics[height=1.25in]{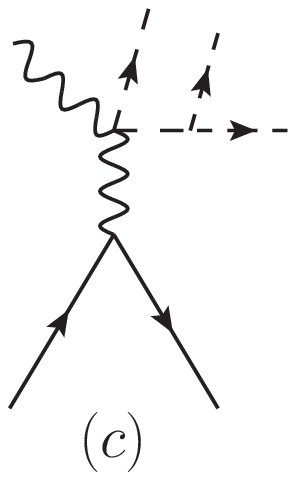}
\end{minipage}\\ [10pt]
\begin{minipage}[t]{0.30\textwidth}
\centering\includegraphics[height=1.25in]{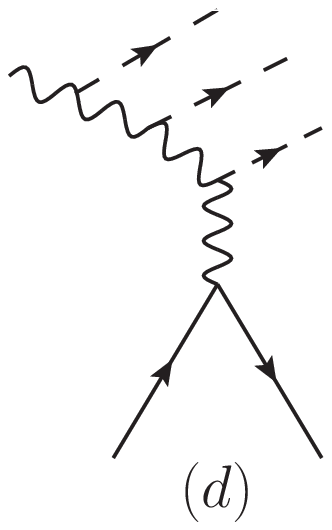}
\end{minipage}%
\begin{minipage}[t]{0.05\textwidth}
\hfil
\end{minipage}%
\begin{minipage}[t]{0.30\textwidth}
\centering\includegraphics[height=1.25in]{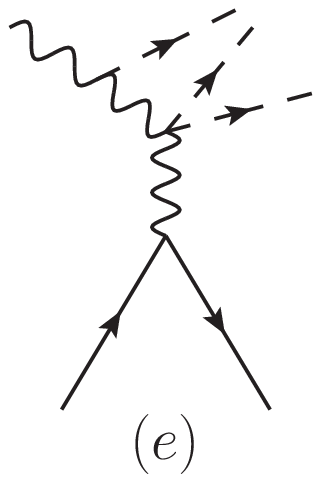}
\end{minipage}%
\begin{minipage}[t]{0.05\textwidth}
\hfil
\end{minipage}%
\begin{minipage}[t]{0.3\textwidth}
\centering\includegraphics[height=1.25in]{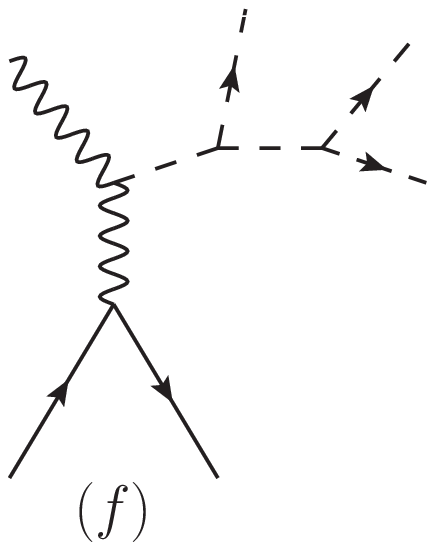}
\end{minipage}%
\caption{Typical diagrams for the process $q\bar{q}\to ZHHH$ are shown. The same set applies to $q\bar{q}\to W^\pm HHH$ . \label{HHHdiag}}
\end{figure}

To get a feeling for the relative strengths of the terms in Eq.\,(\ref{V(H)}) above we consider here the contribution of the subprocesses $q\bar{q}\,\rightarrow\,ZHHH$ to $p\,p\,\rightarrow\,ZHHH+X$ and $q\bar{q}\,\rightarrow\,W^{+}HHH$ to $p\,p\,\rightarrow\,W^{+}HHH+X$. Typical diagrams for this process are shown in Fig.\,(\ref{HHHdiag}).

% Section 2
\section{Contributions from Trilinear and Quartic Couplings}

The matrix element from the Feynman diagrams above has terms of 
the form
\begin{equation}\label{M}
\mathcal{M}\,\sim\,A\kappa_4+B\kappa_3+C+D\kappa_3^2\,,
\end{equation}
where $A$ comes from diagram (a), $B$ from diagrams (b) and (c), $C$ from diagrams (d) and (e), and $D$ from diagram (f). The total cross section is given by
\begin{equation}\label{sigma}
\sigma\,=\,\kappa_4^2\,\sigma_{44}\,+\,\kappa_3^2\,(\sigma_{33}+\sigma_{330})\,+\,\sigma_0\,+\,\kappa_4\kappa_3\sigma_{43}\,+\,\kappa_4\sigma_{40}\,
+\,\kappa_3\sigma_{30}\,+\,\kappa_3^4\sigma_{3333}\,+\,\kappa_4\kappa_3^2\sigma_{433}\,+\,\kappa_3^3\sigma_{333}\,,
\end{equation}
where
\begin{equation}
\begin{array}{lcl}
\sigma_{44}\,\,\,\sim\,|A|^2\,         &\; & \sigma_{33}\,\,\,\,\,\,\sim\,|B|^2\,\\
\sigma_{330}\,\sim\,C\,D^*+C^*\,D\,&\; & \sigma_0\,\,\,\,\,\,\,\,\sim\,|C|^2\,\\
\sigma_{43}\,\,\,\sim\,A\,B^*+A^*\,B\, &\; & \sigma_{40}\,\,\,\,\,\,\sim\,A\,C^*+A^*\,C\,\\
\sigma_{30}\,\,\,\sim\,B\,C^*+B^*\,C\, &\;  & \sigma_{3333}\,\sim\,|D|^2\,\\
\sigma_{433}\,\sim\,A\,D^*+A^*\,D\, &\;& \sigma_{333}\,\,\,\sim\,B\,D^*+B^*\,D
\end{array}
\end{equation}

\begin{figure}[h!]
\centering\includegraphics[height=2.5in]{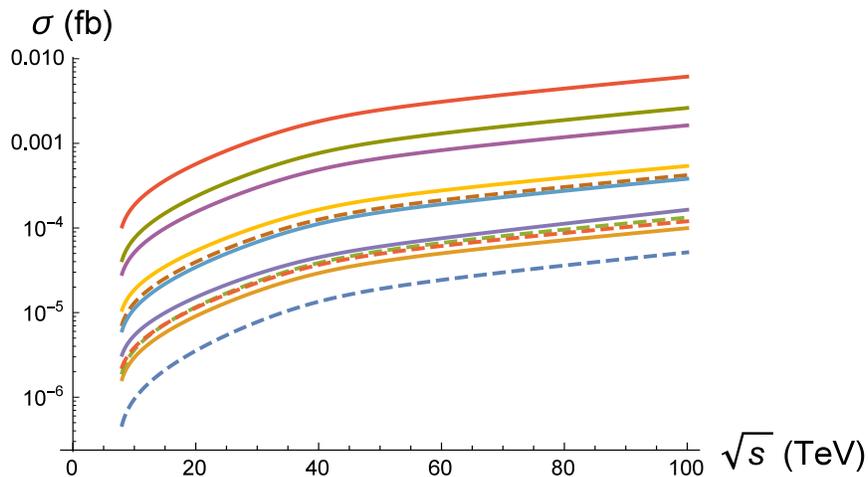}
\caption{The various contributions to the total cross section for $pp\,\rightarrow\,ZHHH+X$ are shown as a function of $\sqrt{s}$. The ordering of the curves corresponds to the ordering of the columns in Table \ref{ZHHH}. $\sigma_{44}$ is the lowest curve and $\sigma_{TOT}$ is the highest. The contributions involving the quartic coupling are indicated by dashed lines. \label{ZHHH_Fig}}
\end{figure}

These separate cross sections for the various terms in Eq.\,(\ref{sigma}), in femtobarns, for several center of mass energies, are given in Table I for $Z$ and Table II for $W^+$. These were derived using CTEQ6L1  distribution functions \cite{CTEQ6L1} with scale $\sqrt{\hat{s}}$. We do not include any contribution from $gq$ or $gg$ initial states.  A $K$ factor of \cite{Han}
\begin{equation}\label{K}
K\,=\,1+\frac{\alpha_s}{2\pi}\pi^2\frac{16}{9}\,\approx\,1.3
\end{equation}
was included where
\begin{equation}
\alpha_s^{-1}\,=\,\frac{1}{0.130}+\frac{21}{12\pi}\log(\frac{\hat{s}}{M_t^2})+\frac{46}{12\pi}\log{\frac{M_t}{M_Z}}\,.
\end{equation}
\begin{sidewaystable}[h]
\begin{center}
\begin{tabular}{|c|c|c|c|c|c|c|c|c|c|c|c|}\hline
\,\,$\sqrt{s}$\,\,&\,\,$\sigma_{44}$\,&\,$\sigma_{3333}$\,&\,$\sigma_{433}$\,&\,$\sigma_{40}$\,&\,$\sigma_{330}$\,&\,$\sigma_{43}$\,&\,$\sigma_{0}$\,&\,$\sigma_{333}$\,
&\,$\sigma_{30}$\,&\,$\sigma_{33}$\,&\,$\sigma_{TOT}$ \\ \hline
 8 & 4.72 $10^{-7}$ & 1.20 $10^{-6}$ & 1.43 $10^{-6}$ & 2.38 $10^{-6}$ & 3.38 $10^{-6}$ & 6.03 $10^{-6}$ & 7.69 $10^{-6}$ & 9.01 $10^{-6}$ &
3.05 $10^{-5}$ & 3.60 $10^{-5}$ & 9.80 $10^{-5}$ \\ 
 13 & 1.57 $10^{-6}$ & 3.61 $10^{-6}$ & 4.47 $10^{-6}$ & 6.94 $10^{-6}$ & 9.31 $10^{-6}$ & 1.80 $10^{-5}$ & 2.32 $10^{-5}$ & 2.55 $10^{-5}$ & 
9.05 $10^{-5}$ & 1.09 $10^{-4}$ & 2.92 $10^{-4}$ \\
 14 & 1.85 $10^{-6}$ & 4.21 $10^{-6}$ & 5.22 $10^{-6}$ & 8.01 $10^{-6}$ & 1.07 $10^{-5}$ & 2.08 $10^{-5}$ & 2.70 $10^{-5}$ & 2.94 $10^{-5}$ & 
1.05 $10^{-4}$ & 1.27 $10^{-4}$ & 3.39 $10^{-4}$ \\
 33 & 9.37 $10^{-6}$ & 1.90 $10^{-5}$ & 2.46 $10^{-5}$ & 3.47 $10^{-5}$ & 4.38 $10^{-5}$ & 9.24 $10^{-5}$ & 1.23 $10^{-4}$ & 1.24 $10^{-4}$ &
4.64 $10^{-4}$ & 5.84 $10^{-4}$ & 1.52 $10^{-3}$ \\
 60 & 2.43 $10^{-5}$ & 4.67 $10^{-5}$ & 6.16 $10^{-5}$ & 8.41 $10^{-5}$ & 1.04 $10^{-4}$ & 2.26 $10^{-4}$ & 3.06 $10^{-4}$ & 2.97 $10^{-4}$ &
1.14 $10^{-3}$ & 1.45 $10^{-3}$ & 3.74 $10^{-3}$ \\
 100 & 5.15 $10^{-5}$ & 9.55 $10^{-5}$ & 1.27 $10^{-4}$ & 1.70 $10^{-4}$ & 2.07 $10^{-4}$ & 4.61 $10^{-4}$ & 6.26 $10^{-4}$ & 5.96 $10^{-4}$ & 
2.31 $10^{-3}$ & 2.97 $10^{-3}$ & 7.62 $10^{-3}$  \\ \hline
\end{tabular}
\end{center}
\caption{Individual contributions to Eq.\,(\ref{sigma}) for $pp\,\rightarrow\,ZHHH+X$.  $\sqrt{s}$ is the center of mass energy in TeV.  All cross sections are in femtobarns. $\sigma_{TOT}$ is the sum of the contributions (the total cross section if $\kappa_3\,=\,\kappa_4\,=\,1$).\label{ZHHH}}
\vspace{.5in}
\begin{center}
\begin{tabular}{|c|c|c|c|c|c|c|c|c|c|c|c|}\hline
\,\,$\sqrt{s}$\,&\,$\sigma_{44}$\,&\,$\sigma_{3333}$\,&\,$\sigma_{433}$\,&\,$\sigma_{40}$\,&\,$\sigma_{330}$\,
&\,$\sigma_{43}$\,&\,$\sigma_{0}$\,&\,$\sigma_{333}$\,&\,$\sigma_{30}$\,&\,$\sigma_{33}$\,&\,$\sigma_{TOT}$ \\ \hline
 8 & 6.58 $10^{-7}$ & 1.63 $10^{-6}$ & 1.96 $10^{-6}$ & 2.27 $10^{-6}$ & 3.17 $10^{-6}$ & 7.26 $10^{-6}$ & 6.14 $10^{-6}$ & 1.07 $10^{-5}$ & 2.87 $10^{-5}$ & 4.16 $10^{-5}$ & 1.04 $10^{-4}$ \\
 13 & 2.03 $10^{-6}$ & 4.53 $10^{-6}$ & 5.65 $10^{-6}$ & 6.00 $10^{-6}$ & 7.96 $10^{-6}$ & 1.99 $10^{-5}$ & 1.70 $10^{-5}$ & 2.78 $10^{-5}$ & 7.79 $10^{-5}$ & 1.17 $10^{-4}$ & 2.85 $10^{-4}$ \\
 14 & 2.36 $10^{-6}$ & 5.19 $10^{-6}$ & 6.52 $10^{-6}$ & 6.58 $10^{-6}$ & 9.02 $10^{-6}$ & 2.28 $10^{-5}$ & 1.96 $10^{-5}$ & 3.16 $10^{-5}$ & 8.93 $10^{-5}$ & 1.34 $10^{-4}$ & 3.27 $10^{-4}$ \\
 33 & 1.08 $10^{-5}$ & 2.12 $10^{-5}$ & 2.78 $10^{-5}$ & 2.66 $10^{-5}$ & 3.34 $10^{-5}$ & 9.15 $10^{-5}$ & 8.08 $10^{-5}$ & 1.21 $10^{-4}$ & 3.55 $10^{-4}$ & 5.54 $10^{-4}$ & 1.32 $10^{-3}$ \\
 60 & 2.66 $10^{-5}$ & 5.01 $10^{-5}$ & 6.65 $10^{-5}$ & 6.13 $10^{-5}$ & 7.56 $10^{-5}$ & 2.13 $10^{-4}$ & 1.91 $10^{-4}$ & 2.77 $10^{-4}$ & 8.29 $10^{-4}$ & 1.31 $10^{-3}$ & 3.10 $10^{-3}$ \\
 100 & 5.43 $10^{-5}$ & 9.95 $10^{-5}$ & 1.33 $10^{-4}$ & 1.20 $10^{-4}$ & 1.64 $10^{-4}$ & 4.21 $10^{-4}$ & 3.81 $10^{-4}$ & 5.40 $10^{-4}$ & 1.63 $10^{-3}$ & 2.62 $10^{-3}$ & 6.14 $10^{-3}$ \\ \hline
\end{tabular}
\end{center}
\caption{Same as Table \ref{ZHHH} except for $pp\,\rightarrow\,W^{+}HHH+X$.\label{WHHH}}
\end{sidewaystable}
For the process $pp\,\rightarrow\,ZHHH+X$, the contents of Table \ref{ZHHH} are illustrated in Fig. \ref{ZHHH_Fig}. The figure for $pp\,\rightarrow\,W^{+}HHH+X$ is similar.
%\newpage

The amplitude $C$ in Eq.\,(\ref{M}) comes from the $ZZH$ and $ZZHH$
couplings (diagrams (d) and (e)).  Superficially these diagrams grow
faster with energy than the diagrams that involve Higgs
propagators. However, the largest energy behavior cancels between the
diagrams with only $ZHH$ couplings and those that involve a $ZZH$ and
a $ZZHH$ coupling. Explicitly implementing this cancellation of the
large energy behavior seems essential for the calculation of
$\sigma_0$;  depending on the phase space integral to find the
cancellation does not work for large center of mass energies. A
similar high energy behavior occurs in the $W^+$ cross section and
requires the same analytic cancellation. 
A detailed description of how this cancellation occurs is given 
in the next section.

% Section 3
%\appendix
\section{Cancellation of the leading high energy behavior}

If we label the momenta as 
\begin{equation}
q(p_1)+\bar{q}(p_2)\,\rightarrow\,H(k_1)+H(k_2)+H(k_3)+Z(P)
\end{equation}
then the matrix element for $C$ can be written
\begin{equation}\label{M0}
M\,\sim\,\bar{v}(p_2)\gamma_{\mu}(g_V-\gamma_5)u(p_1)X^{\mu\lambda}\epsilon_{\lambda}(P)
\end{equation}
The spinor factor goes as $E^1$ at large energy $E$.  $X^{\mu\lambda}$ has two or three $Z$ propagators depending on the diagram. The propagator which couples to the spinor factor goes as $E^{-2}$ because the momentum in the $p^{\mu}p^{\nu}/M_Z^2$ term is $p_1+p_2$ which is zero when dotted into spinor factor.  The other one or two propagators do not have this cancellation and thus go as $E^{0}$. The $Z$ polarization vector can be longitudinal and thus go as $E^{1}$.  So these diagrams go as $E^{0}$ for large $E$.  The diagrams for contributions other than $C$ go as $E^{-2}$ or faster because they have Higgs propagators.

To see this $E^{0}$ behavior cancel we need to write out $X^{\mu\lambda}$
\begin{eqnarray}
X^{\mu\lambda}\,&=&\,\frac{1}{2}A_1^{\mu\rho}(B_{2\rho}^{\lambda}+B_{3\rho}^{\lambda}+g_{\rho}^{\lambda})
+\frac{1}{2}A_2^{\mu\rho}(B_{1\rho}^{\lambda}+B_{3\rho}^{\lambda}+g_{\rho}^{\lambda})
+\frac{1}{2}A_3^{\mu\rho}(B_{1\rho}^{\lambda}+B_{2\rho}^{\lambda}+g_{\rho}^{\lambda})\,\nonumber\\
&+&\,\frac{1}{2}(A_1^{\mu\rho}+A_2^{\mu\rho}+g^{\mu\rho})B^{\lambda}_{3\rho}
+\frac{1}{2}(A_2^{\mu\rho}+A_3^{\mu\rho}+g^{\mu\rho})B^{\lambda}_{1\rho}
+\frac{1}{2}(A_1^{\mu\rho}+A_3^{\mu\rho}+g^{\mu\rho})B^{\lambda}_{2\rho}
\end{eqnarray}  
where
\begin{eqnarray}
A_i^{\mu\lambda}\,&=&\,C_i(M_Z^2g^{\mu\lambda}+k_i^{\mu}Q_i^{\lambda})\,\,\,\,\,\,\,\,{\rm no\,\,sum\,\,on\,\,i} \\
B_i^{\mu\lambda}\,&=&\,D_i(M_Z^2g^{\mu\lambda}-R_i^{\mu}k_i^{\lambda})\,\,\,\,\,\,\,\,{\rm no\,\,sum\,\,on\,\,i}
\end{eqnarray}
for $i=1,2,3$ with 
\begin{eqnarray}
Q_i^{\mu}\,&=&\,p_1^{\mu}+p_2^{\mu}-k_i^{\mu} \\
R_i^{\mu}\,&=&\,P^{\mu}+k_i^{\mu}
\end{eqnarray}
and
\begin{eqnarray}
C_i\,=\,\frac{1}{Q_i^2-M_Z^2} \\
D_i\,=\,\frac{1}{R_i^2-M_Z^2}
\end{eqnarray}

The large $E$ behavior comes from the $P^\mu P^\nu/M_Z^2$ part of the sum over $Z$ polarizations, so replace the polarization vector $\epsilon_{\lambda}(P)$ by $P_\lambda$ and dot $P_{\lambda}$ into $X^{\mu\lambda}$. Then use 
\begin{equation}
D_i^{-1} = 2P\Dot k_i+\frac{1}{2}m_H^2
\end{equation}
to eliminate $P\Dot k_i$ factors in favor of mass factors or the cancellation of $D_i$ terms. The remaining large $E$ terms will occur in the combination  $C_i(p_1+p_2-k_i)^2$, which can be replaced by $M_Z^2$ and terms that vanish when contracted with the lepton factor. In particular if we define 
\begin{equation}
F_i^{\rho}\,=\,D_i[M_Z^2P^{\rho}+\frac{1}{2}M_H^2(P+k_i)^{\rho}]
\end{equation}
then
\begin{eqnarray}\label{XMU}
X^{\mu\rho}P_{\rho}\,&=&\,\frac{1}{2}A_i^{\mu\rho}(F_{j\rho}+F_{k\rho})+\frac{1}{2}[A_i^{\mu\rho}+A_j^{\mu\rho}+g^{\mu\rho}]F_{k\rho} \nonumber \\
                     &\equiv&\,X^{\mu}
\end{eqnarray}
where 
\begin{equation}
i,j,k\,=\,(1,2,3)\,+\,(2,3,1)\,+\,(3,1,2)
\end{equation}
and $F_i^{\rho}$ is smaller than $B_i^{\rho\lambda}P_{\lambda}$ by two factors of mass rather than momenta. (The terms with the additional factors of momenta are proportional to $P^{\mu}+k_1^{\mu}+k_2^{\mu}+k_3^{\mu}\,=\,p_1^{\mu}+p_2^{\mu}$ dotted into the spinor factor.)

If we call the square of the spinor factor in Eq.\,(\ref{M0}), summed over spins, $L_{\mu\nu}$, then the square of the matrix element for $\sigma_0$, including the transverse polarizations of the $Z$, is
\begin{equation}
\sum_{pol}|M|^2\,\sim\,L_{\mu\nu}X^{\mu\lambda}X^{\nu\eta}(-g_{\lambda\eta})\,+\,L_{\mu\nu}X^{\mu}X^{\nu}/M_Z^2
\end{equation}
where $X^{\mu}$, defined in Eq.\,(\ref{XMU}) above, can be simplified to
\begin{equation}
X^{\mu}\,=\,A_1^{\mu\rho}(F_{2\rho}+F_{3\rho})+A_2^{\mu\rho}(F_{1\rho}+F_{3\rho}) +A_3^{\mu\rho}(F_{1\rho}+F_{2\rho})+\frac{1}{2}g^{\mu\rho}(F_{1\rho}+F_{2\rho}+F_{3\rho})\,.
\end{equation}
By explicitly implementing this cancellation, the integration over phase space is well behaved for all  beam energies.

% Section 4
\section{Results and Conclusions}

\begin{figure}[h!]
\begin{minipage}[t]{0.45\textwidth}
\centering\includegraphics[height=2.0in]{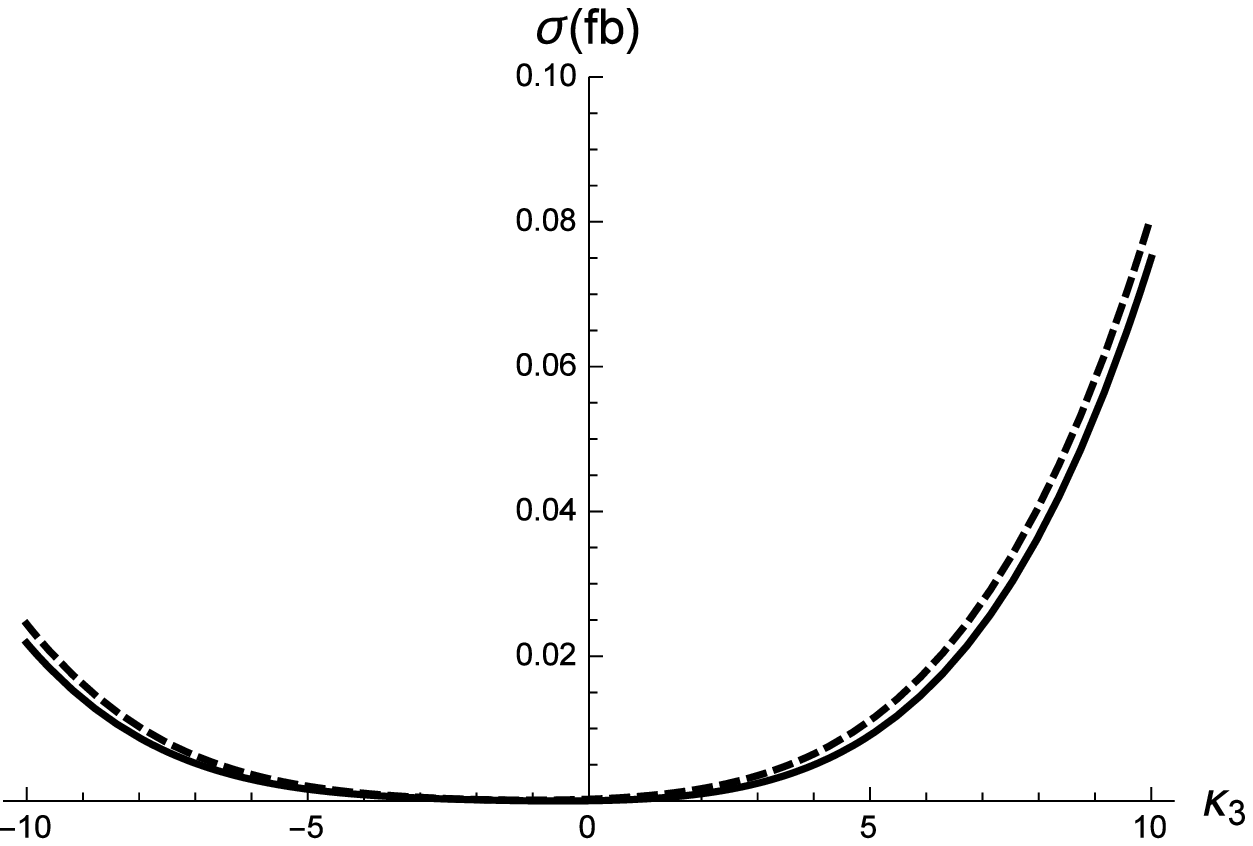}
\caption{The cross section for $pp\,\rightarrow\,ZHHH+X$ from $q\bar{q}\,\rightarrow\,ZHHH$ for $\sqrt{s}=13$ TeV is shown as a function of $\kappa_3$ for $\kappa_4=1$ (solid line) and $\kappa_4=10$ (dashed line).\label{ZHHH13}}
\end{minipage}%
\begin{minipage}[t]{0.1\textwidth}
\hfil
\end{minipage}%
\begin{minipage}[t]{0.45\textwidth}
\centering\includegraphics[height=2.0in]{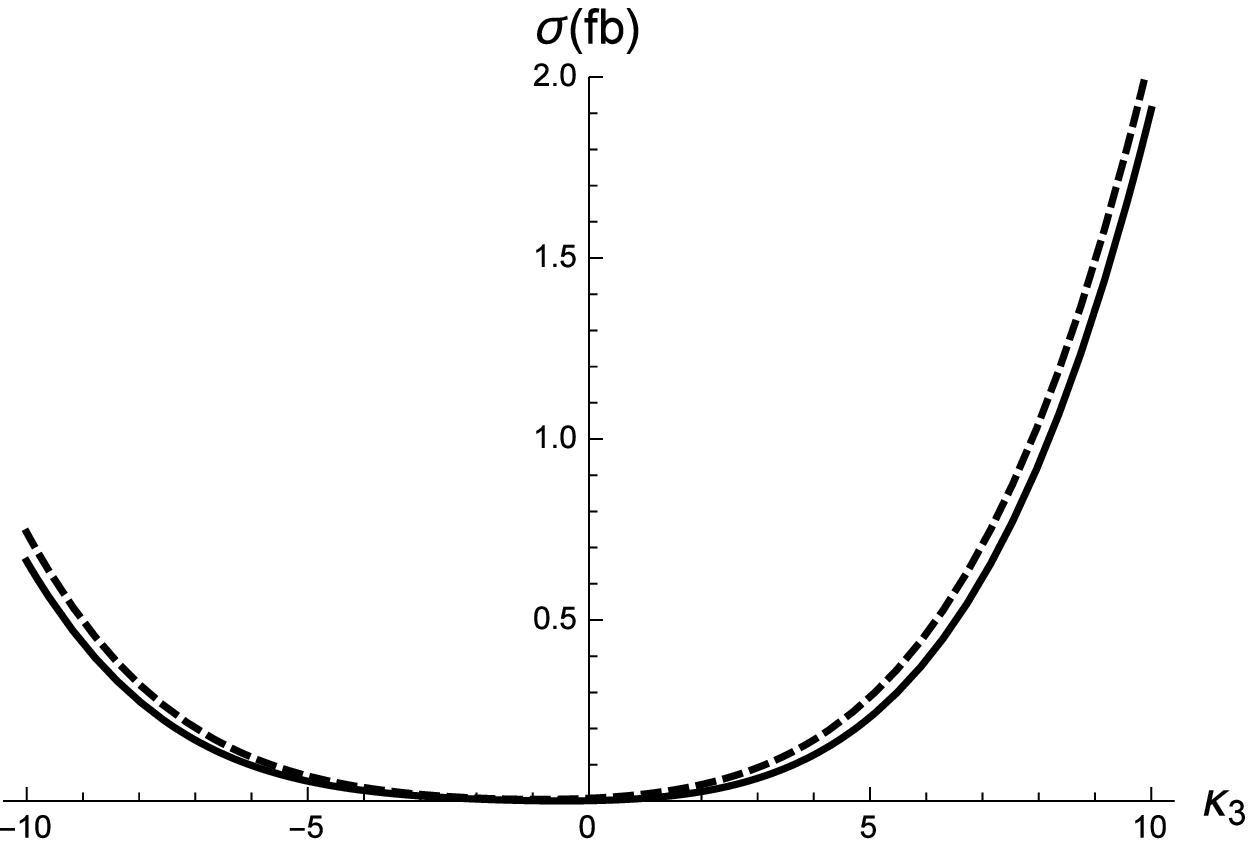}
\caption{The cross section for $pp\,\rightarrow\,ZHHH+X$ from $q\bar{q}\,\rightarrow\,ZHHH$ for $\sqrt{s}=100$ TeV is shown as a function of $\kappa_3$ for $\kappa_4=1$ (solid line) and $\kappa_4=10$ (dashed line).\label{ZHHH100}}
\end{minipage}%
\end{figure}

The Tables show that the coefficients of $\kappa_4$ in the cross
section Eq.\,(\ref{sigma}) are small which makes a value for
$\kappa_4$ almost impossible to determine independent of the value of
$\kappa_3$. For example Figs.\,\ref{ZHHH13} and \ref{ZHHH100} show the
cross section for the $Z$ process with $\sqrt{s}\,=\,13$ and
$\sqrt{s}\,=\,100$ as a function of $\kappa_3$ for two values of
$\kappa_4$. 
\begin{figure}[h!]
\begin{minipage}[t]{0.45\textwidth}
\centering\includegraphics[height=2.0in]{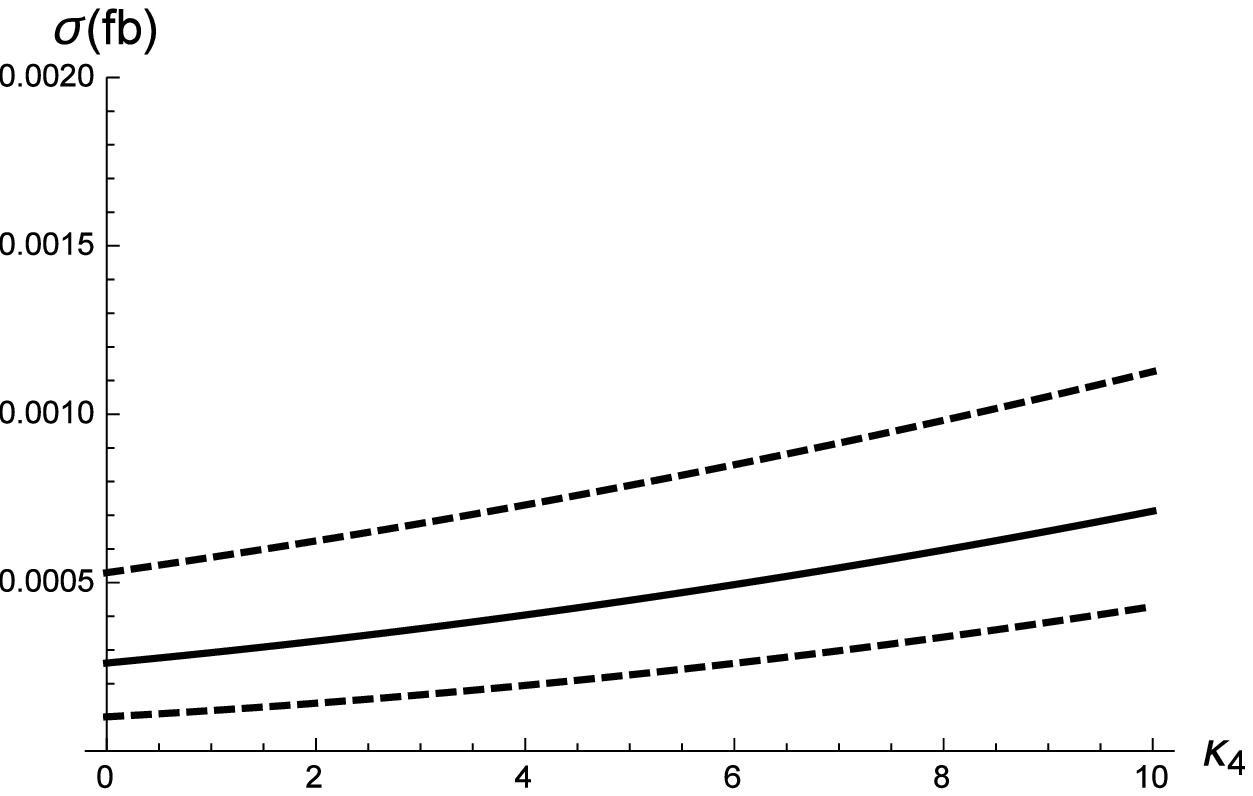}
\caption{The variation of the cross section for $pp\,\rightarrow\,ZHHH+X$ from $q\bar{q}\,\rightarrow\,ZHHH$ for $\sqrt{s}=13$ TeV is shown for $\kappa_3=1$ (solid line) and the dashed band $0.5\leq\kappa_3\leq 1.5$ as a function of $\kappa_4$.\label{k413}}
\end{minipage}%
\begin{minipage}[t]{0.1\textwidth}
\hfil
\end{minipage}%
\begin{minipage}[t]{0.45\textwidth}
\centering\includegraphics[height=2.0in]{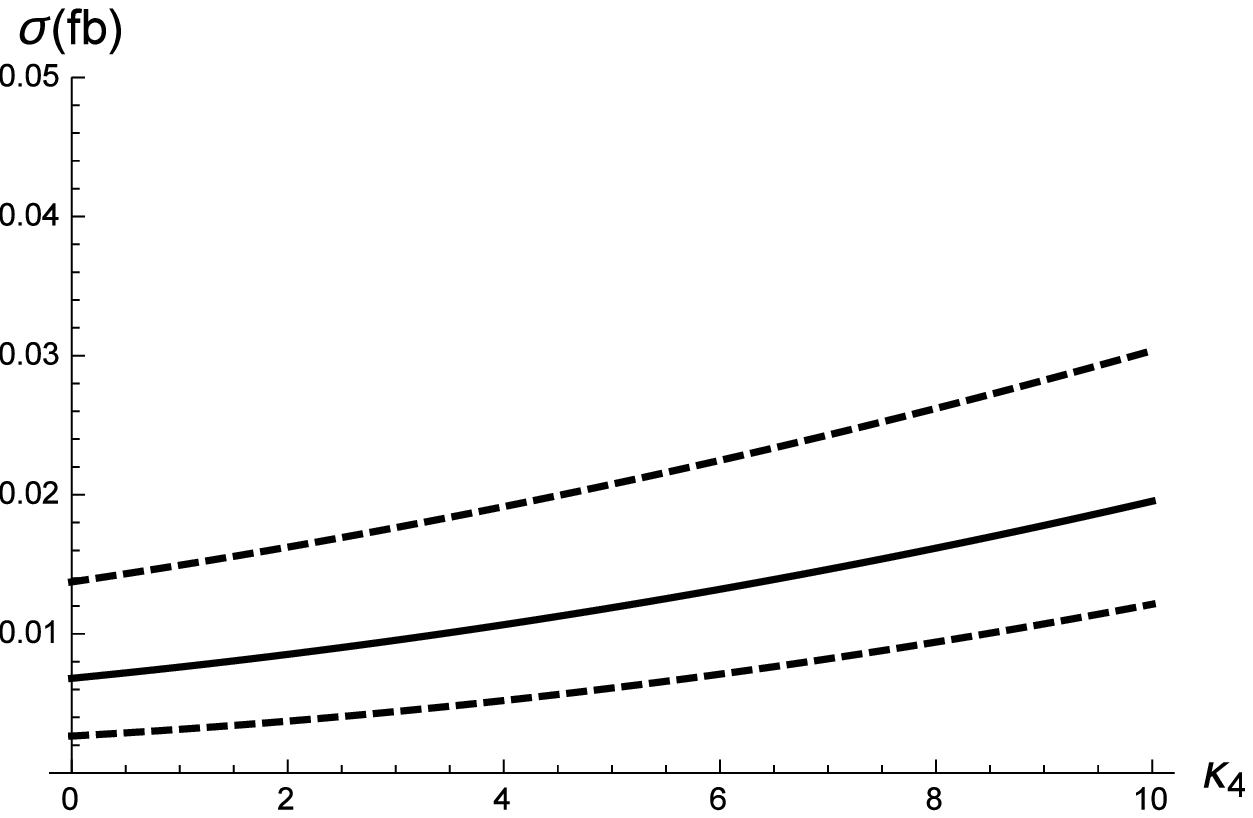}
\caption{The variation of the cross section for $pp\,\rightarrow\,ZHHH+X$ from $q\bar{q}\,\rightarrow\,ZHHH$ for $\sqrt{s}=100$ TeV is shown for $\kappa_3=1$ (solid line) and the dashed band $0.5\leq\kappa_3\leq 1.5$ as a function of $\kappa_4$.\label{k4100}}
\end{minipage}%
\end{figure}
At $\sqrt{s}\,=\,13$ TeV with $\kappa_3\,=\,1$ the difference in the
cross section between $\kappa_4\,=\,1$ and $\kappa_4\,=\,10$ is
$4.2\times10^{-4}$ fb. For $\sqrt{s}\,=\,100$ TeV the same difference
is $1.2\times10^{-2}$ fb and if $\kappa_3\,=\,10$ the difference is
still less than $0.16$ fb. Figures \ref{k413} and \ref{k4100}
illustrate this further by fixing $\kappa_3$ near $1$ and varying
$\kappa_4$ to find that the cross sections change by only small
fractions of a femtobarn. For the $W^{+}$ process the contributions 
that include the quartic Higgs coupling are again too small 
to measure $\kappa_4$ or even to determine if it is nonzero. 
This is illustrated in Figs.\,\ref{WHHH13} and \ref{WHHH100}.

\begin{figure}[h!]
\begin{minipage}[t]{0.45\textwidth}
\centering\includegraphics[height=2.0in]{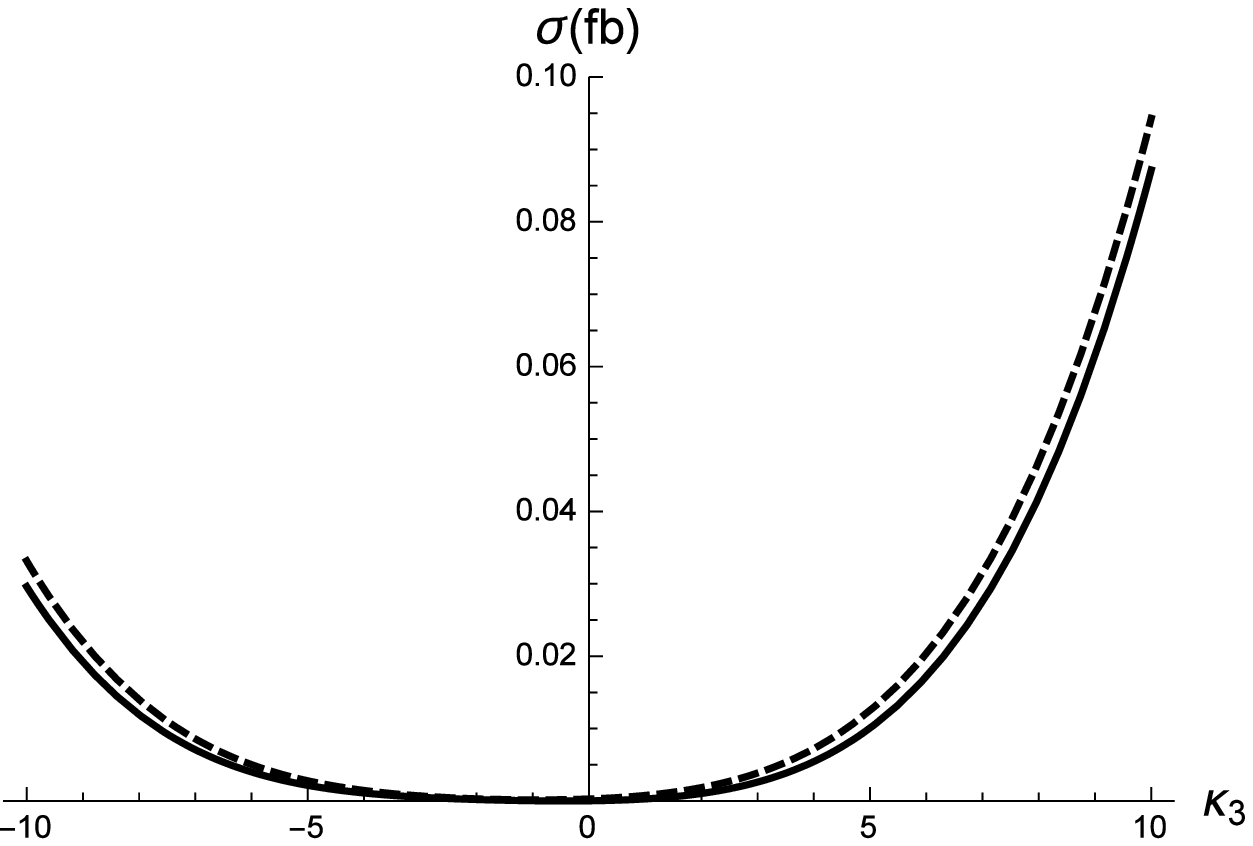}
\caption{The cross section for $pp\,\rightarrow\,W^+HHH+X$ from $q\bar{q}\,\rightarrow\,W^+HHH$ for $\sqrt{s}=13$ TeV is shown as a function of $\kappa_3$ for $\kappa_4=1$ (solid line) and $\kappa_4=10$ (dashed line).\label{WHHH13}}
\end{minipage}%
\begin{minipage}[t]{0.1\textwidth}
\hfil
\end{minipage}%
\begin{minipage}[t]{0.45\textwidth}
\centering\includegraphics[height=2.0in]{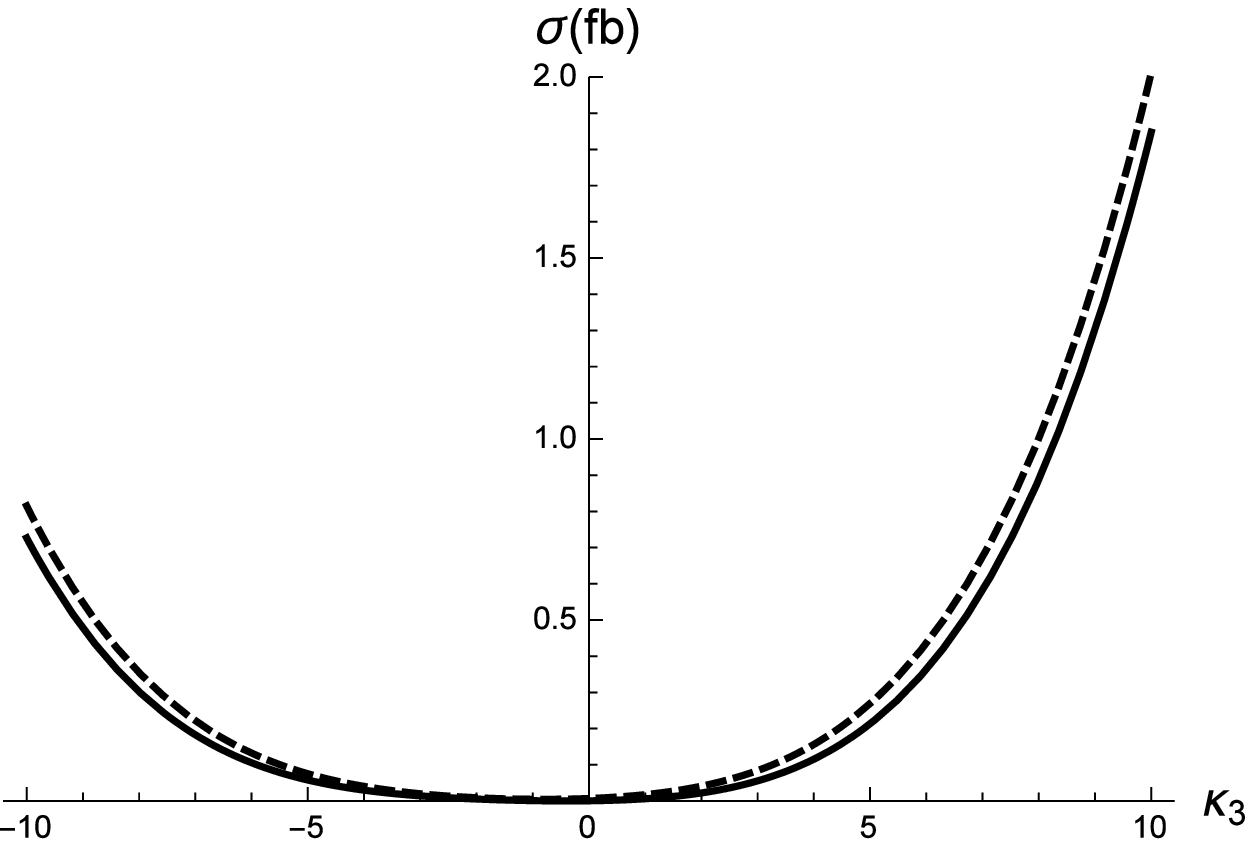}
\caption{The cross section for $pp\,\rightarrow\,W^+HHH+X$ from $q\bar{q}\,\rightarrow\,W^+HHH$ for $\sqrt{s}=100$ TeV is shown as a function of $\kappa_3$ for $\kappa_4=1$ (solid line) and $\kappa_4=10$ (dashed line).\label{WHHH100}}
\end{minipage}%
\end{figure}

The total cross section for the $W^{-}$ process is smaller than that for $W^{+}$ by a factor of $2.66,\, 2.24,\, 2.19,\, 1.75,\, 1.57,\, 1.47$ for $\sqrt{s}\,=\,8,\, 13,\, 14,\, 33,\, 60,\, 100$ TeV.   The ratios of the individual cross sections (eg., $\sigma_{44}$) vary from these numbers by less than $10\%$.

The parameter $\kappa_3$ can be determined from processes with two Higgs bosons in the final state. For example, the subprocess $gg\,\rightarrow\,HH$ obviously depends on the three Higgs coupling as does $gg\,\rightarrow\,t\bar{t}HH$ \protect\cite{DKW,Glover,Daw,Dol,Bag,Goe,Barr,Arh,deFlorian,Hes,Liu,deFl,Fred,dawson,DKR,Cao1,Cao2}. Processes with three Higgs bosons in the final state are necessary to determine $\kappa_4$. We show that the processes considered here are not
sufficient at any energy to even verify the existence of a four Higgs coupling. This is most obvious from Figure 2 where the coefficients of $\kappa_4$ (dashed lines) are very small compared to most of the other partial cross sections.  In general the problem of determining $\kappa_4$ will be very difficult. Similar conclusions have been reached by Binoth, Karg, Kauer, and R\"uckl \cite{BKKR} and others \cite{PS,Chen} for the gluon fusion process $gg\to HHH$.

\noindent{\bf Acknowledgements}\\
D.~A.~D. was supported in part by the U.~S. Department of Energy under Award No.DE-FG02-12ER41830, C.~K. was supported in part by the U.~S. Department of Energy under Award No.DE-FG02-13ER41979 and W.~W.~R. was supported in part by the National Science Foundation under Grant No. PHY 1068020.

%
% References
%

\end{document}